\documentclass[final,3p,sort&compress,twocolumn]{elsarticle}

\usepackage[pagewise]{lineno}
\modulolinenumbers[1]
\usepackage{hyperref}
\usepackage{svg}
\hypersetup{colorlinks=true,pdfborder=0 0 0}
\usepackage{subcaption}
\usepackage{amsmath,bm,siunitx,amssymb,caption,booktabs}
\usepackage{ulem}
\usepackage{xspace}

\usepackage{textcomp,gensymb}
\usepackage{cleveref}
\usepackage{ulem}
\usepackage{xcolor}
\usepackage{fancyhdr}

\usepackage{etoolbox}
\makeatletter
\patchcmd{\ps@pprintTitle}
  {Preprint submitted}
  {Postprint of article submitted}
  {}{}
\makeatother

\journal{Scripta Materialia}

\begin{document}

\begin{frontmatter}

\title{FIP-GNN: Graph neural networks for scalable prediction of grain-level fatigue indicator parameters}

\address[snu-mse] {Department of Materials Science and Engineering \& RIAM, Seoul National University, 1 Gwanak-ro, Gwanak-gu, Seoul 08826, Korea}
\address[az-mse]{Department of Materials Science and Engineering, University of Arizona, Tucson, AZ 85721, USA}
\address[az-am]{Graduate Interdisciplinary Program in Applied Mathematics, University of Arizona, Tucson, AZ 85721, USA}

\author[snu-mse]{Gyu-Jang Sim}
\author[snu-mse]{Myoung-Gyu Lee}
\author[az-mse,az-am]{Marat I. Latypov\corref{cor1}}
\cortext[cor1]{corresponding author}
\ead{latmarat@arizona.edu}

\begin{abstract}

High-cycle fatigue is a critical performance metric of structural alloys for many applications. The high cost, time, and labor involved in experimental fatigue testing call for efficient and accurate computer models of fatigue life. We present FIP-GNN -- a graph neural network for polycrystals that (i) predicts fatigue indicator parameters as grain-level inelastic responses to cyclic loading quantifying the local driving force for crack initiation and (ii) generalizes these predictions to large microstructure volume elements with grain populations well beyond those used in training. These advances can make significant contributions to statistically rigorous and computationally efficient modeling of high-cycle fatigue -- a long-standing challenge in the field. \\

\noindent Note: \textcolor{red}{this is an author-generated postprint} of the article by Sim et al.\ \href{https://doi.org/10.1016/j.scriptamat.2024.116407}{published} in {\it{Scripta Mater.}} (2025). \\ DOI: 10.1016/j.scriptamat.2024.116407

\end{abstract}


\end{frontmatter}

\pagestyle{fancy}
\fancyhf{}
\fancyhead[LO]{Postprint of \href{https://doi.org/10.1016/j.scriptamat.2024.116407}{Sim et al., Scr Mater (2025) 116407}}

Understanding and predicting fatigue are crucial in designing and qualifying structural alloys. Determining alloy lifespan in high cycle fatigue (HCF) is challenging due to its variability \cite{miao2009crystallographic} and the time-consuming cycling tests needed for crack initiation. The cost, time, and labor of experimental HCF testing highlight the need for efficient and accurate models.

\begin{figure*}[!ht]
\centering
\includegraphics[width=\linewidth]{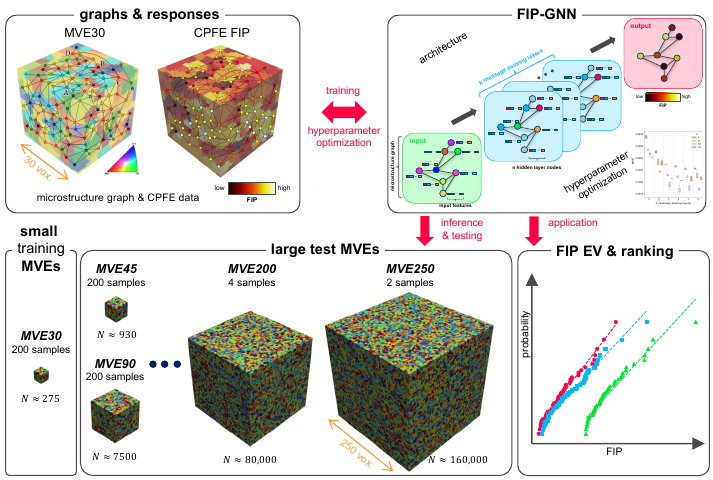}
\caption{Pictorial overview of this study, including training and optimization of GNNs with small MVEs (MVE30 set), GNN architecture, large MVEs used for testing the predictive capabilities of GNNs for modeling grain-level FIPs, and predicting FIP EV distributions as the application for trained GNNs.  MVE sets (MVE30, MVE45, \ldots, MVE250) are labeled according to the number of voxels (``vox.'') along one side of the cube MVEs. Only one example MVE is visualized for each set, total number of MVEs (``samples'') and approximate number of grains, $N$, for each MVE size are also provided. }

\label{fig:overview}
\end{figure*}

One approach to modeling HCF aims to compute the driving force for crack initiation as crack initiation accounts for most of the HCF life in polycrystalline alloys \cite{McDowell1989}. The driving force is often quantified by a fatigue indicator parameter (FIP) calculated in microstructure-based simulations, e.g., crystal plasticity finite element (CPFE) method \cite{McDowell2010}. Calculation of FIPs 
that best predict crack initiation has been subject to intensive research \cite{Rovinelli_2017}, leading to the introduction of various FIPs  \cite{McDowell2010,Barenblatt1987,Riemelmoser1997,Rolfe1977,Fatemi1988}. The Fatemi--Socie FIP is widely used due to its low mesh sensitivity \cite{Castelluccio2012} and its accessibility within continuum CPFE simulations without explicit crack tip implementation. Utilizing the Fatemi--Socie FIP as a metric of the driving force for crack initiation, Przybyla et al.\ \cite{Przybyla2010} developed a methodology of evaluating and ranking microstructures in terms of their HCF resistance. Their approach uses statistical analysis of FIP values in polycrystalline microstructure volume elements (MVEs). However, the method faces challenges because statistical analysis of most interest for the HCF resistance focuses on extreme value (EV) distributions that require MVEs prohibitively large for CPFE simulations. For an Al 7075-T6 alloy, MVEs containing even $10^6$ grains were reported insufficiently representative for FIP EV distributions, which is already impractically expensive -- a CPFE simulation on one MVE with only 1000 grains requires over 100 CPU hours \cite{Yaghoobi2021}. Given the high computational cost and the need in large grain populations for EV distributions, researchers adopted the concept of statistical volume elements (SVEs), where multiple volume elements represent one microstructure in lieu of one large MVE \cite{GU2023118715,priddy2017strategies,Paulson2018}. SVEs allow estimating FIP EV distributions from large grain populations in aggregate, though it is still computationally expensive due to the need in CPFE simulations on multiple volume elements for each microstructure. 

The high computational cost of the CPFE method for statistically rigorous modeling of HCF emphasizes the need in new efficient approaches. Machine learning is a promising route for reducing the computational cost of modeling microstructure--property relationships. A machine learning strategy extensively explored in literature is the development of surrogate models that reproduce the results of computationally demanding direct numerical simulations such as CPFE at a fraction of their computational cost. Quite a few surrogate models have been reported to date \cite{latypov2019materials,Paulson2017,Paulson2018,Pandey2021,Ibragimova2022,vlassis2020geometric,pagan2022graph,Sadeghpour2022,hestroffer2023graph,Thomas2023,Karimi2023,hu2024anisognn, Dai2021,Yang2022,Dai2023,HU2024104017,HANSEN2024108019,QIAN2024104046}. 

\begin{figure*}[t]
\centering
\includegraphics[width=\linewidth]{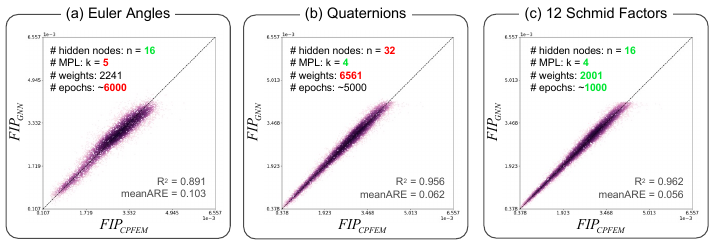}
\caption{Parity plots comparing grain-level FIPs predicted by the GNN vs.\ those from CPFE simulations for 20 microstructures from the MVE30-VAL set. The parity plots are shown for three GNNs with 
optimized hyperparameters ($n$ and $k$) based on (a) Euler angles, (b) quaternions, (c) 12 Schmid factors as graph node features. The corresponding MeanARE and $R^2$ as well as the number of learnable weights and number of epochs to reach a plateau in the validation loss are also shown.}
\label{fig:node_feature}
\end{figure*}

The surrogate models primarily differ in their approach to quantitative description of microstructure. Reported descriptions include spatial correlations \cite{latypov2019materials,Paulson2017,Paulson2018}, latent representation by convolutional neural networks learned during training \cite{Frankel2020,Pandey2021,Ibragimova2022,hu2022learning}, and graphs \cite{vlassis2020geometric,pagan2022graph,Sadeghpour2022,hestroffer2023graph,Thomas2023,Karimi2023,hu2024anisognn, Dai2021,Yang2022,Dai2023,HU2024104017}. Graph representation of polycrystals and use of graph neural networks (GNNs) have been gaining momentum, as graphs capture grain structure and connectivity in polycrystals without requiring full 3D voxel data unlike other descriptions. 
 GNNs have been successful in modeling various properties of polycrystals, both mechanical \cite{vlassis2020geometric,pagan2022graph,Sadeghpour2022,hestroffer2023graph,Thomas2023,Karimi2023,hu2024anisognn,HU2024104017} and non-mechanical \cite{Dai2021,Yang2022,Dai2023}. Vlassis et al.\ \cite{vlassis2020geometric} developed GNN models to predict the elastic energy in hyperelastic materials. Pagan et al.\ \cite{pagan2022graph} trained GNNs to predict grain-average stress in Ni and Ti alloys using CPFE and experimental data. Sadeghpour et al.\ \cite{Sadeghpour2022} modeled the yield strength of 316L steel, while Hestroffer et al.\ \cite{hestroffer2023graph} developed GNNs for the yield strength and elastic modulus of $\alpha$-Ti. Hu and Latypov \cite{hu2024anisognn} generalized GNNs ability to predict anisotopic properties in arbitrary loading directions.  Thomas et al.\ \cite{Thomas2023} developed GNNs to predict fatigue based on experimental data on surface protrusions. Hu et al.\ \cite{HU2024104017} combined a GNN with recurrent neural networks to predict grain orientation and stress history in polycrystals. These studies demonstrated the potential of GNNs for modeling overall properties and grain-level responses in polycrystals and render GNNs promising for computationally efficient FIP prediction. While machine learning of FIPs have been recently pursued -- e.g., Hansen et al.\ \cite{HANSEN2024108019} used symbolic regression for learning interpretable dependence of FIPs on microstructure; Qian et al.\ \cite{QIAN2024104046} utilized physics-informed neural network to predict energy density-based FIP in through silicon via copper; Paulson et al.\ \cite{Paulson2018} developed reduced-order models based on spatial statistics and principal component analysis -- GNNs have yet to be explored for modeling FIPs.

In this study, we demonstrate the ability of GNNs to model grain-level FIPs as non-linear micromechanical responses of polycrystals under cyclic loading. We show that GNNs can tackle the challenge of modeling FIPs in large MVEs with numerous grains at a modest computational cost. To this end, we trained GNNs using publicly available dataset on Al 7075-T6 alloy \cite{stopka_2022} featuring CPFE simulation results by Stopka, Yaghoobi et al.\ \cite{Stopka2022,Yaghoobi2023}. The dataset includes computer-generated MVEs of various sizes ($30^3$ to to $250^3$ voxels, see \Cref{fig:overview}) and grain counts (\SI{275}{} to \SI{160000}{}) and their simulated micromechanical responses, including spatially-resolved FIPs. For each MVE, we created a microstructure graph whose nodes represent grains and whose edges link adjacent grains, thereby representing grain boundaries. 
We incorporated different grain properties as node features, including (i) Euler angles and (ii) quaternions -- both describing grain orientations -- as well as (iii) Schmid factors relative to the loading direction. As a node response for learning and inference, we calculated grain-level Fatemi--Socie FIPs by averaging maximum values from CPFE integration points in a grain. We trained and optimized the GNNs with the three feature choices using a set of 200 smallest MVEs (MVE30) and then tested the trained models on large MVEs. \Cref{fig:overview} shows a sketch of FIP-GNN, further details of the GNN model, data, training and hyperparameter optimization are described in the Supplementary Material.

\begin{figure*}[t]
\centering
\includegraphics[width=0.8\linewidth]{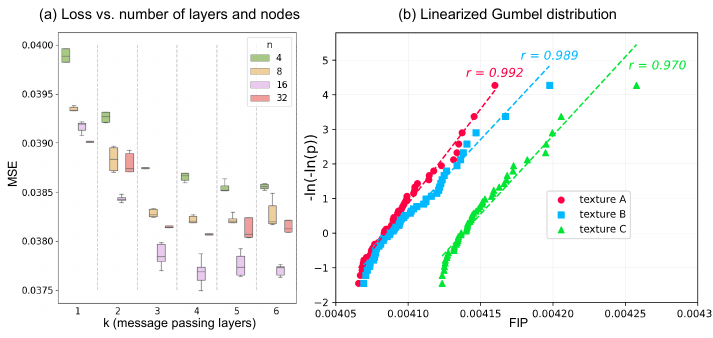}
\caption{(a) Results of hyperparameter optimization for the GNN model trained with 12 Schmid factors as features shown in terms of the mean squared error (MSE) for a validation set (MVE30-VAL), which was obtained for different combinations of the number of nodes in the hidden layer, $n$, and the number of message passing layers, $k$. (b) Gumbel extreme value distribution of FIP values predicted by the optimized GNN for microstructures with three crystallographic textures: cubic texture (A, red circles), nearly uniform texture (B, blue squares), and rolled (C, green triangles), $r$ represents Pearson correlation coefficient of each EV FIP distribution.}
\label{fig:results}
\end{figure*}

We first compare GNNs trained with different node features: Euler angles, quaternions, and Schmid factors. We split the MVE30 set to 180 training and 20 validation MVEs to optimize GNN hyperparameters and evaluate the impact of the feature choice on the GNN performance. All GNNs capture grain-level FIPs with $R^2>0.89$ and mean absolute relative error (MeanARE) within \SI{10}{\percent} (\Cref{fig:node_feature}). The GNN using Euler angles as features had the lowest accuracy (\Cref{fig:node_feature}a), while those using quaternions and Schmid factors showed comparably superior accuracy. However, quaternion-based GNN (\Cref{fig:node_feature}b) required over three times more learnable weights than the Schmid factor-based GNN (\Cref{fig:node_feature}c) to achieve similar accuracy due to the higher optimal number of hidden layer nodes obtained in hyperparameter optimization (see $n$ in \Cref{fig:node_feature}b,c). 
The GNN using Schmid factors also learned fastest reaching validation error plateau in \SI{1000}{} epochs compared to the \SI{5000}{} to \SI{6000}{} epoch range for orientation-based GNNs. Given this combination of accuracy, size, and speed, we focus on GNNs using Schmid factors for the rest of the study.

\begin{figure*}[!ht]
\centering
\includegraphics[width=0.9\linewidth]{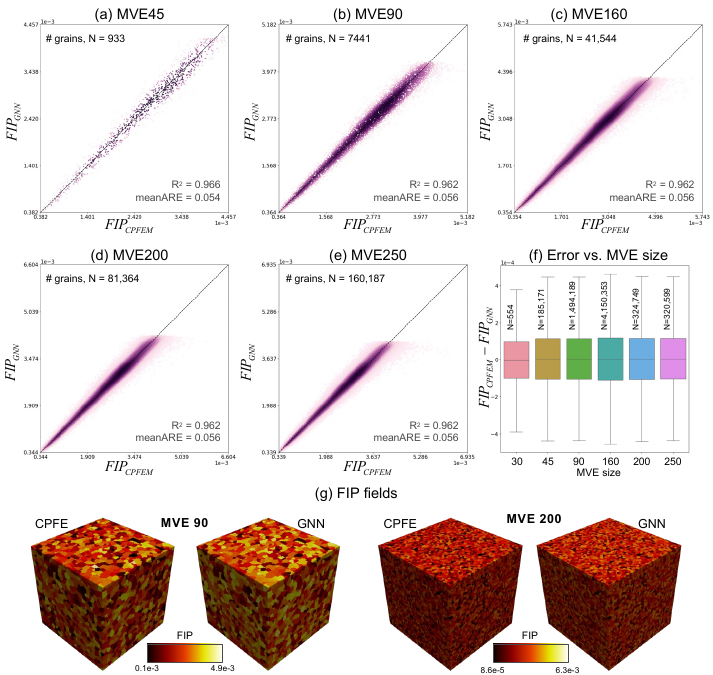}
\caption{Evaluation of the capability of the GNN trained on 200 microstructures of the MVE30 set to predict FIPs in larger MVEs shown as (a-e) parity plots comparing GNN FIPs with ground truth values from CPFE simulations for individual MVEs of increasing size; (f) box plot showing the distribution of the difference in predicted FIPs from ground truth values; (g) grain-level FIP fields predicted by GNN vs.\ FIP fields from CPFE simulations. }
\label{fig:generalization}
\end{figure*}

We next evaluate the GNN's ability to generalize FIP predictions to large MVEs with grain populations well beyond those used in training. We trained the best GNN identified above with Schmid factors as features and optimized hyperparameters $n$ and $k$ (\Cref{fig:results}a) on the full MVE30 set and tested its accuracy on MVEs of increasing sizes from $45^3$ to $250^3$ voxels. Predictions for one MVE from each set show that the GNN accurately predicts grain-level FIPs for large MVEs with significantly larger grain populations than seen during training (\Cref{fig:generalization}). The GNN consistently achieves accuracy of $R^2>0.96$ for five MVEs representing all studied sizes. 
Extending this analysis to all MVEs in each set, we summarize the results in a box plot of error distributions (\Cref{fig:generalization}f). The difference between GNN-predicted and CPFE values remains close to zero across all MVE sizes.
The error range does not correlate with the MVE size, which again attests to the GNNs ability to predict FIPs for MVEs with large grain populations with consistent accuracy. 
Finally, FIP fields predicted by the GNN visually match the CPFE fields as shown in representative example MVEs (\Cref{fig:generalization}g). 

To evaluate the ability of FIP-GNN to efficiently rank microstructures, we analyzed EV distributions of FIP values. We calculated EV distributions for microstructures with three textures distinct from the training set: (A) cubic texture, (B) weak, nearly uniform texture, and (C) rolled texture. Using a public dataset  \cite{mc174Stopka2021} for Al 7075-T6, we compared FIP-GNN predictions with EV distributions and fatigue performance rankings from high-fidelity CPFE simulations. 
We converted three $90^3$ MVEs into graphs, obtained FIPs at each grain with the trained GNN, and extracted 50 highest FIP values for each microstructure.  We linearized the distribution of the 50 values using the Gumbel distribution, following prior analyses of FIP EV distributions from CPFE simulations \cite{stopka_texture_article,Przybyla2010}. \Cref{fig:results}b shows that the 50 highest FIP values predicted by GNN fit the Gumbel distribution (Pearson correlation $r>0.97$) in all three cases. Furthermore, the GNN-based EV distributions correctly rank the microstructures: texture C performs best and texture A the worst, consistent with CPFE simulations by Stopka et al.\ \cite{Stopka2022,stopka_texture_article}. We emphasize that the GNN correctly ranks microstructures with textures distinct from those in the training set, which included only MVEs with uniform textures (i.e., random grain orientations). This analysis confirms the ability of FIP-GNN to capture EV distributions and qualitatively rank individual microstructures in the order consistent with high-fidelity simulations even for microstructures distinct from the training dataset. This is despite some variability in GNN-predicted FIPs at higher magnitudes (\Cref{fig:generalization}b--e).

The presented GNN approach enables not only FIP predictions and microstructure ranking but also insights into micromechanics of polycrystals relevant to FIPs and HCF. Using 12 Schmid factors per grain as node features resulted in the most accurate GNN model (\Cref{fig:node_feature}) with minimal learnable weights. To explore if all 12 Schmid factors are necessary, we tested a GNN model focusing on Schmid factors of only three most active slip systems per grain. With three highest Schmid factors as features, the GNN model achieved $R^2=0.956$ for the MVE30-VAL set. This accuracy is only marginally lower than the full 12-factor model and nearly identical to the quaternion-based GNN  (\Cref{fig:node_feature}b-c). This surprisingly good accuracy of a GNN that uses only three Schmid factors suggests that the driving force for crack initiation quantified by FIPs is associated with slip activity on very few dominant systems. This is consistent with arguments in literature that few slip systems get activated in each grain during deformation of polycrystals \cite{LEE2010925}. 
Further tests showed that even one or two Schmid factors for most active systems still allowed FIP predictions with $R^2>0.8$. In addition, the comparable accuracy of the GNN with three Schmid factors and the quaternion-based GNN indicates that Schmid factors provide better accuracy {\it{not}} because they offer more features. This clearly follows from the identical accuracy of the GNN with three Schmid factors as three features ($R^2=0.956$) compared to the GNN with quaternions as four features ($R^2=0.956$). Schmid factors lead to more accurate GNNs at the same or even smaller number of features than orientations can be attributed to more direct relevance to the response of interest (FIPs) and better compatibility with aggregation functions in GNNs \cite{hu2024anisognn}. While Schmid factors sufficed in this study of uniaxial cycling of polycrystals, elements of full Schmid tensors (as in \cite{hu2024anisognn}) can serve as features for modeling HCF under more complex, multi-axial loading conditions.

We can gain further insight into the role of local microstructure in FIP and crack initiation from the GNN architecture optimized in this study. Our hyperparameter tuning identified the optimal number of message passing layers, $k$. Each message passing layer aggregates features from the nearest-neighbor nodes in the graph \cite{Hamilton2017} so that stacking $k$ layers aggregates features from neighbors $k$ edges apart. In the context of polycrystal micromechanics, $k$ signifies the range of the local neighborhood that affects the response (FIP) in a given grain. We found $k=4$ layers optimal for predicting grain-level FIPs (see \Cref{fig:results}a), meaning Schmid factors up to fourth-order nearest neighbors are necessary for accurate FIP predictions. This data-driven result is consistent with explicit studies of the impact of grain neighborhood on FIPs using physics-based CPFE simulations. Stopka et al.\ \cite{Stopka2022} found that the FIP value in a ``hot-spot'' grain is sensitive to the orientations of the nearest neighbor grains of up to fourth order for the same alloy studied here. Similarly, our results show that accounting for features from grains beyond four edges away (i.e., $k>4$) did not improve the accuracy of FIP predictions (\Cref{fig:results}a).

In conclusion, we developed GNNs for polycrystals that, for the first time (to our knowledge), can (i) predict FIPs as grain-level responses to cyclic loading quantifying the local driving force for crack initiation, and (ii) generalize these predictions to MVEs with grain populations significantly larger than MVEs used in training. These advances contribute to statistically rigorous and computationally efficient modeling of HCF -- a long standing challenge in the field. The computational gains of presented GNNs are two-fold: (i) GNNs serve as orders of magnitude faster surrogates to CPFE simulations of FIPs, and (ii) GNNs can obtain FIPs and their distributions in large MVEs out of reach for direct CPFE simulations. Indeed, inference with trained FIP-GNN took under \SI{200}{\milli\second} for the largest MVEs  in this study (MVE250) and just \SI{2.5}{\milli\second} for MVE30 samples on a consumer-grade workstation. In contrast, direct CPFE simulations of the same FIPs after cyclic loading require over 100 hours for one MVE90 sample on a single CPU \cite{Yaghoobi2021}. Moreover, unlike CPFE simulations \cite{5744437}, GNNs exhibit $O(N)$ complexity, meaning their computational cost increases linearly with the number of grains, $N$, in an MVE. This opens opportunities for rapid calculation of statistically significant FIP EV distributions in microstructures for their ranking in terms of HCF resistance and design for superior HCF life. We confirmed that FIP-GNN can correctly rank microstructures by FIP EV distributions, even for MVEs with textures distinct from those used in training. Finally, we demonstrated that the GNN approach provides insights into the micromechanics of polycrystals during HCF loading consistent with prior physics-based CPFE modeling. 

\section*{Code availability}

The codes created as part of this study are available on GitHub at \href{https://github.com/materials-informatics-az/FIP-GNN}{\url{https://github.com/materials-informatics-az/FIP-GNN}}.

\section*{Acknowledgements}

GJS and MGL acknowledge the support from NRF of Korea (grant No.\ 2022R1A2C2009315) and from the Ministry of Science and ICT (grant No.\ 2022M317A4072293). The authors thank Dr.\ Krzysztof Stopka (Purdue University) for help with processing the MVE/FIP dataset used in this study. GJS further thanks Yunju Jang for assistance with some of the illustrations.

\section*{Supplementary material}

This supplementary material provides the details on the (i) data used for training and testing FIP-GNNs; (ii) microstructure representation with graphs; (iii) GNN architecture design; (iv) GNN training, hyper-parameter tuning, and validation; (v) error metrics.

\subsection*{Data}
\label{input_data}
    
For training GNNs in this study, we made use of a dataset published by Stopka and Yaghoobi \cite{stopka_2022}, which contains an ensemble of polycrystalline MVEs and the micromechanical data from CPFE simulations carried out on these MVEs. The MVEs represent 3D polycrystalline aggregates of an Al 7075-T6 alloy with predominantly equiaxed microstructure and uniform texture \cite{Yaghoobi2023}. The grain size of the MVEs follows a lognormal distribution with an average equivalent grain diameter of \SI{14}{\micro\meter} and its standard deviation of \SI{2}{\micro\meter}. The crystallographic orientation of each grain in the dataset is described by Euler angles in the ZXZ convention. The dataset contains six subsets of MVEs differentiated by their size that cover a broad spectrum of grain populations. The subsets correspond to six MVE sizes with \SI{30}{}, \SI{45}{}, \SI{90}{}, \SI{160}{}, \SI{200}{}, and \SI{250}{} voxels along each of the three principal axes. These sizes correspond to the average grain counts per MVE of about \SI{280}{}, \SI{930}{}, \SI{7500}{}, \SI{41000}{}, \SI{80000}{}, and \SI{160000}{} grains, respectively. We refer to these MVE subsets as MVE30, MVE45, MVE90, MVE160, MVE200, and MVE250, where the number indicates the MVE size in terms of the voxel number in one direction. In terms of the micromechanical data relevant to this study, the dataset includes Socie--Fatemi FIP values calculated with CPFE simulations. The simulations captured tension-compression loading of the MVEs along the $x$ axis for two cycles (sufficient for convergence of FIP \cite{Stopka2022}) and strain amplitude of 0.7\% in a completely reversed manner ($R_\varepsilon=-1$). From these simulations, we adopted the grain-average FIP values as the specific grain-level responses to be modeled with GNNs. FIP used in this study is the Fatemi--Socie FIP defined as \cite{Przybyla2010}

\begin{equation}\label{eqn:fip}
\text{FIP}_\alpha=\frac{\Delta \gamma^{\alpha}_{p}}{2} \left[ 1 + K\frac{\sigma^{\alpha}_{n}}{\sigma_{y}}\right],
\end{equation}

\noindent where $\Delta \gamma^{\alpha}_{p}$ is the range of plastic shear strain and $\sigma^{\alpha}_{n}$ is the maximum normal stress on the $\alpha^\text{th}$ slip system, $K$ is a parameter quantifying the influence of normal stress (set to 10), and $\sigma_{y}$ is the macroscopic yield strength.

\subsection*{Microstructure graph, node features and response variables}

\label{microstructure_graph}
    \begin{figure}[!ht]
    \centering
    \includegraphics[width=0.66\linewidth]{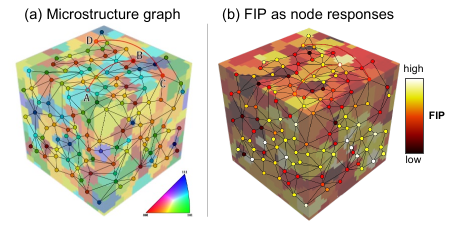}
    \caption{ (a) Conversion from MVE to graph. Each grain is represented as a node, and each grain boundary is represented as an edge linking adjacent grains. To account for periodicity of the MVE, node $A$ is ensured to be connected to node $B$, and node $C$ is connected to node $D$. (b) Illustration of the assignment of grain-average FIP from CPFE fields to graph nodes.}
    \label{fig:mve2graph}
    \end{figure}

To leverage GNNs for modeling FIPs, we create a microstructure graph for each MVE in the dataset. In the microstructure graph, grains are represented by graph nodes, while grain boundaries are represented by graph edges that link nodes corresponding to spatially adjacent grains (\Cref{fig:mve2graph}a). The MVEs in the dataset are periodic so that the construction of the graphs needs to account for the microstructure periodicity. Specifically, nodes that represent grains on one side of the MVE need to be connected to the nodes representing grain neighbors on the opposite side ($A$--$B$ and $C$--$D$ node pairs in \Cref{fig:mve2graph}a for example). Furthermore, when a grain is cut by a side of the MVE and thus its part appears on the opposite side, the grain needs to be represented by a single node in the graph. To satisfy the continuity of the microstructure in the graphs of periodic MVEs, we adopt a simple method that duplicates the given MVE and appends its 26 dummy copies surrounding the original MVE at its six faces, 12 edges and eight corners. Once we have the MVE augmented with dummy copies, we identify grain neighbors using the DREAM.3D software \cite{Groeber2014}. From the grain neighbor data, we finally construct a graph with the NetworkX library \cite{osti_960616}. 

We then introduce grain-level properties as node features into the constructed microstructure graphs. We tested two orientation representations: (i) Euler angles, and (ii) quaternions. The MVE dataset that we adopted had triplets of Euler angles in radians as the raw orientation data, which can be readily used as three-element feature vectors. For quaternions as features, we converted the Euler angles into rotation matrices from which we then calculated the four-element quaternion vectors \cite{rowenhorst2015consistent}. We further considered Schmid factors of individual grains in respect to the loading direction. Schmid factors quantify resolved shear stresses on each slip system that ultimately drive the dislocation glide in individual grains \cite{Schmid}. For each node in the microstructure graph, we thus assigned 12 Schmid factors calculated for the 12 slip systems in aluminum in respect to the loading direction in the CPFE simulations.

In addition to ``input'' features, we introduced grain-level FIP values as a response variable for inference at each node (\Cref{fig:mve2graph}b). Following the calculation method by Przyblyla et al.\ \cite{Przybyla2010}, we determined the grain-wise FIP by averaging the maximum FIP values over all integration points of each grain. At each integration point, the maximum FIP is found among all slip systems calculated in CPFE \cite{stopka_2022} according to \Cref{eqn:fip}. The grain-average FIP values are the node properties modeled by the FIP-GNNs developed in this study. 

\subsection*{GNN architecture design, optimization, and training}
\label{gnn_model_architecture}

The GNN architecture considered in this study contained message passing layers and a final output layer as the key components of the architecture. A message passing layer aggregates graph node features from first-order nearest neighbors \cite{DBLP:journals/corr/GilmerSRVD17} using an aggregation or convolution function. Stacking $k$ message passing layers results in aggregation of features from $k^\text{th}$ order neighbors. We considered multiple convolution methods, including SAGE \cite{Hamilton2017}, GCN \cite{Kipf2017}, and GIN \cite{xu2019powerful}. Among these methods, SAGE convolution demonstrated significantly better performance in terms of validation error, and was therefore chosen. SAGE convolution includes hidden layers with $n$ neural nodes that process features of each graph node. We treated the number of message passing layers, $k$, the number of neural nodes in the hidden layer, $n$, as well as learning parameters (optimizer, rate, decay, number of warm-up steps) as the main hyperparameters that we optimized to our data.

\begin{figure}[!ht]
\centering
\includegraphics[width=\linewidth]{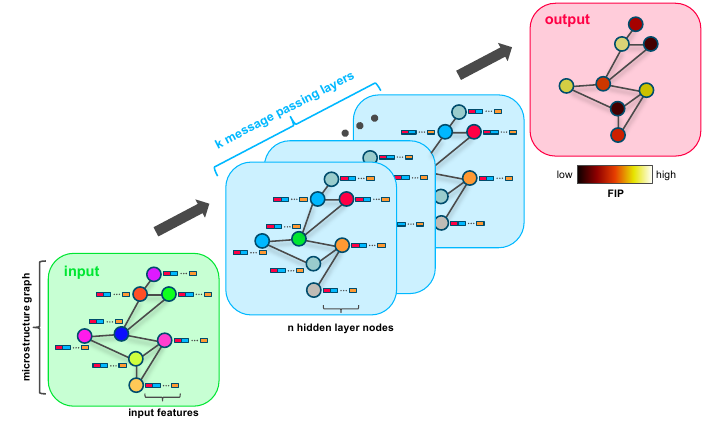}
\caption{Simplified illustration of the architecture of FIP-GNN comprised including the input graph, $k$ message passing layers (featuring SAGE convolution with $n$ neural nodes in the hidden layer), and output graph with FIPs as node responses.}
\label{fig:architecture}
\end{figure}

To ensure generalization capability of GNNs to large MVEs, we trained the models and optimized their hyperparameters using only a set of 200 MVEs of the smallest size -- MVE30. For hyperparameter optimization, we randomly split the set of 200 MVE30 into training (MVE30-TRAIN) and validation (MVE30-VAL) sets in the $90:10$ ratio. Using validation error as the optimization metric, we found that the convolution method, number of message passing layers, $k$, and the number of hidden layers, $n$, had the biggest impact on the GNN training outcome. SAGE convolution was found the best among the tested three methods, while the optimal combinations of the $k$ and $n$ parameters depended on the selected node features (orientations vs. Schmid factors) as summarized in Figure 2 of the main text. Figure 3a in the main text further shows the effect of $k$ and $n$ on the mean squared error for the validation set (MVE30-VAL).

\subsection*{Error metrics}
\label{error_estimates}

This work uses three error metrics for analysis: mean squared error (MSE), mean absolute relative error (MeanARE), and coefficient of determination ($R^2$). These metrics are defined as follows.

\begin{subequations}
\begin{equation}\label{eqn:MSE}
\text{MSE}=\frac{1}{n}\sum_{i=1}^{n}(y_i-\hat{y}_i)^2,
\end{equation}

\begin{equation}\label{eqn:MeanARE}
\text{MeanARE}=\frac{1}{n}\sum_{i=1}^{n}\frac{|y_i-\hat{y}_i|}{|y_i|},
\end{equation}

\begin{equation}\label{eqn:R2}
R^2=1-\frac{\sum_{i=1}^{n}(y_i-\hat{y}_i)^2}{
\sum_{i=1}^{n}(y_i-\bar{y}_i)^2},
\end{equation}
\end{subequations}

\noindent where \(y_i\) is the ground truth value, \(\hat{y}_i\) is the predicted value, \(\bar{y}_i\) is the mean of the ground truth values, and \(n\) is the total number of data points.

\bibliography{references}
\end{document}